\documentclass[final]{svjour2}
\usepackage{graphicx}
\usepackage{rotating}
\usepackage{amssymb}
\usepackage{mathptmx}
\usepackage[numbers]{natbib}
\makeatletter
\journalname{Journal of Low Temperature Physics}

\bibpunct{}{}{,}{s}{}{,}

\begin{document}

\newcommand{\hdblarrow}{H\makebox[0.9ex][l]{$\downdownarrows$}-}
\title{Background suppression in massive TeO$_2$ bolometers with Neganov-Luke amplified light detectors}

\author{L.~Pattavina$^{1*}$\and
N.~Casali$^2$\and
L.~Dumoulin$^3$\and
A.~Giuliani$^3$\and
M.~Mancuso$^3$\and
P.~de~Marcillac$^3$\and
S.~Marnieros$^3$\and
S.S.~Nagorny$^6$\and
C.~Nones$^4$\and
E.~Olivieri$^3$\and
L.~Pagnanini$^6$\and
S.~Pirro$^1$\and
D.~Poda$^3$\and
C.~Rusconi$^5$\and
K.~Sch\"aeffner$^1$\and
M.~Tenconi$^3$
}

\institute{$^1$INFN Ð Laboratori Nazionali del Gran Sasso, I-67100 Assergi (AQ), Italy\\
$^*$luca.pattavina@lngs.infn.it\\
$^2$INFN Ð Sezione di Roma I, I-00185 Roma, Italy\\
$^3$CSNSM, Centre de Sciences NuclŽaires et de Sciences de la Mati\`ere,  CNRS/IN2P3, Universit\`e Paris-Sud, 91405 Orsay, France\\
$^4$CEA, Centre d`Etudes Saclay, IRFU, 91191 Gif-Sur-Yvette Cedex, France\\
$^5$INFN Ð Sezione di Milano-Bicocca I, I-20126 Milano, Italy\\
$^6$Gran Sasso Science Institute, I-67100 LÕAquila, Italy}

\maketitle

\begin{abstract}

Bolometric detectors are excellent devices for the investigation of neutrinoless double-beta decay (0$\nu\beta\beta$). The observation of such decay would demonstrate the violation of lepton number, and at the same time it would necessarily imply that neutrinos are Majorana particles.
The sensitivity of cryogenic detectors based on TeO$_2$ is strongly limited by the $\alpha$-background in the region of interest for the 0$\nu\beta\beta$ of $^{130}$Te. It has been demonstrated that particle identification in TeO$_2$ bolometers is possible measuring the Cherenkov light produced by particle interactions. However, the discrimination efficiency is low and an event-by-event identification with NTD-based light detectors has to be demonstrated.
We will discuss the performance of a highly-sensitive light detector exploiting the Neganov-Luke effect for signal amplification. The detector, being operated with NTD-thermistor and coupled to a 750~g TeO$_2$ crystal, shows the ability for an event-by-event identification of electron/gamma and $\alpha$ particles. The obtained results demonstrate the possibility to enhance the sensitivity of TeO$_2$-based 0$\nu\beta\beta$ experiment to an unprecedented level.


\end{abstract}

\section{Introduction}
Double-beta decay is a weak nuclear process, where two neutrons transform into two protons and it is followed by the emission of two electrons and two neutrinos. This process is known as two neutrino double-beta decay (2$\nu\beta\beta$). It is also postulated that, in the final state there are just two electrons and nothing else. This decay is known as neutrinoless double-beta decay (0$\nu\beta\beta$) and it manifestly violates the lepton number conservation. The rate of such rare process is strictly related to the effective neutrino Majorana mass. The observation and measurement of the half-life of 0$\nu\beta\beta$ combined with the neutrino oscillation parameters would allow us to constrain the neutrino mass and hierarchy.

In the standard interpretation of 0$\nu\beta\beta$ in terms of neutrino mass mechanism, two challenges have to be faced by experimentalists. The first one consists in approaching and then covering the inverted hierarchy region of the neutrino mass pattern, and the second and ultimate goal is to explore the direct hierarchy region. CUPID (Cuore Upgrade with Particle IDentification)\cite{CUPID-0}, a proposed bolometric 0$\nu\beta\beta$ experiment, try to face the first of these two challenges~\cite{CUPID-1} developing innovative techniques.

\section{Background suppression in bolometric experiment}
The possibility to study rare processes is strongly influenced by the background level in the region of interest. This challenge is presently faced by different bolometric experiments involved in the search for extremely weak signals in overwhelming backgrounds, such as 0$\nu\beta\beta$ decay.

The main background sources are produced by natural radioactivity. Bolometers are fully active detectors, thus the pursuit for an ultra-low background experiment is an effortful task to be fulfilled. The main contaminants in such detectors are $^{232}$Th and $^{238}$U decay chain products, which emit $\alpha$ and $\beta$/$\gamma$ particles with energy close to the ROI for 0$\nu\beta\beta$ decay investigation. Nevertheless, the benefits that this technique offers are extraordinary and still not completely exploited. A key advantage of bolometers is the wide choice of detector materials, and if the detector compound is properly selected, we may have a detector made of the $\beta\beta$ decay source.

In this context, the CUORE experiment~\cite{CUORE}, which will operate the first ton-scale bolometric detector, aims at investigating the 0$\nu\beta\beta$ decay of $^{130}$Te using TeO$_2$ crystals as absorbers. The expected sensitivity that CUORE will reach after 5~y of data taking, operating 988 TeO$_2$ bolometers, will be 9.5$\cdot$10$^{25}$~y at 90\% C.L. with a background index in ROI (Q$_{\beta\beta}=$2.5~MeV) of 10$^{-2}$~c/keV/kg/y and an energy resolution of 5~keV.

As already demonstrated by CUORE-0~\cite{CUORE0}, CUORE sensitivity will be mainly limited by $\alpha$ decays occurring on the detector surfaces~\cite{sticking}, namely the copper structure~\cite{TTT} and the crystals~\cite{CCVR}.

In this work we show an effective active method for identifying and rejecting $\alpha$ interactions in the detector and thus enhancing the sensitivity for a next-generation ton-scale experiment based on TeO$_2$ bolometers. The read-out of Cherenkov light emitted in TeO$_2$ crystals using Neganov-Luke amplified light detectors allow to discriminate $\alpha$ particles from $\beta$/$\gamma$ events on an event-by-event basis.

\section{Cherenkov light emission in TeO$_2$ crystals}
In 2010, Tabarelli de Fatis~\cite{Tabarelli} proposed Cherenkov emission from $\beta$ rays in bolometric crystals as a viable alternative to scintillation. The read-out of Cherenkov photons makes tagging of $\beta$/$\gamma$ events in TeO$_2$ crystals possible, thanks to the optical properties of the compound. The energy threshold for electron for producing Cherenkov photons is at about 50~keV, while $\alpha$ particles require much greater energy, at a level of 400~MeV, well above natural radioactivity. If a dedicated cryogenic light detector sensitive to the single Cherenkov photon is operated together with a TeO$_2$ absorber, a full-background rejection of $\alpha$ particles is achievable. At 2.5~MeV ($^{130}$Te $\beta\beta$ Q-value), Cherenkov photons have energies in a range between 2 eV (about 600 nm) and 3.5 eV (about 350 nm).

The signal produced by $^{130}$Te 0$\nu\beta\beta$ decay is expected to produce a light signal of about 300~eV~\cite{Tabarelli}. In literature, various Cherenkov light measurements with TeO$_2$ crystals can be found; the measured light signals in the ROI are always smaller than the expected ones.
\begin{table*}[htdp]
\begin{center}
\caption{State of the art of Cherenkov light measurement with large mass TeO$_2$ cystals.} 
\begin{tabular}{cccccc}
\hline\noalign{\smallskip}
TeO$_2$ mass & LD thermal & $\beta$/$\gamma$ detected light & RMS energy resolution & ref.\\ 
\noalign{}\noalign{}
[g] & sensor & @ ROI [eV] &  @ 0 eV [eV] & ~\\
\noalign{\smallskip}\hline\noalign{\smallskip}
750  & Ge-NTD & 101 & 72 & ~\cite{Casali}\\
\noalign{\smallskip}\hline\noalign{\smallskip}
285  & W-TES & 129 & 23 & ~\cite{Schaeffner}\\
\noalign{\smallskip}\hline\noalign{\smallskip}
117 & Ge-NTD & 195 & 97 & ~\cite{TeO2_small} \\
\noalign{\smallskip}\hline
\end{tabular}
\label{Tab1} 
\end{center}
\end{table*}
In Table~\ref{Tab1}, the most relevant Cherenkov light measurements with massive TeO$_2$ absorbers are listed. Looking at the values in the table, we can remark an anti-correlation between the mass of the crystal and the detected light. We discern that most of the light does not escape from the crystal due to self-trapping and internal surface reflections. A detailed description of Cherenkov light propagation and light collection efficiency in TeO$_2$ crystal can be found in~\cite{MC_light}. Nevertheless, the light signal is very weak, and Ge-Neutron Transmutation Doped (Ge-NTD) based light detectors (LD)~\cite{LD_performance} used for $\beta\beta$ applications, are not yet optimized for such low energy threshold. An efficient particle discrimination relying on the detection of such tiny light signals is tempting. Currently, many efforts are spent for enhancing the performance of bolometric LD. The final goal is to develop wide area detectors sensitive to the single photon.

\section{Neganov-Luke amplified light detectors}

In the field of thermal detectors for rare events investigations, two types of thermal sensors have been developed: W-based Transition Edge Sensors (W-TES) and Ge-NTD thermistors. While the first class ensures excellent performance in terms of energy resolution (RMS at 0~keV is about 10~eV~\cite{CRESST}) but low production reproducibility, the second does not exhibit performance at the same level (RMS at 0~keV is about 80~eV~\cite{LD_performance}), but did prove to be highly reproducible~\cite{CUORE0}. For this reason, highly performing LD based on Ge-NTD thermal sensors are very promising devices.

So far, standard Ge-NTD based LDs consist of highly pure germanium disks operated as bolometers, these show performance good enough for the detection of light emitted by scintillating bolometers used in 0$\nu\beta\beta$ applications~\cite{LUCIFER}, but not enough for an effective Cherenkov light measurement with TeO$_2$ detectors.

Taking advantage of the Neganov-Luke effect~\cite{Luke1,Luke2}, the detector performance, in terms of energy threshold,
can be enhanced when operating Ge-NTD based LDs. During particle interaction in the LD (e.g. Cherenkov photons), phonons and charged carriers are produced. If an electrical field is applied on the detector, electron-hole pairs will drift across the field while generating additional phonons, resulting in an amplification of the thermal signal. In first approximation, the gain linearly scales with the voltage across the detector:
\begin{equation}
G = 1 + \frac{e\cdot V}{\epsilon}
\end{equation}
where, $e$ is the electron charge, $V$ the applied voltage and $\epsilon$ the energy needed to create an electron-hole pair.\newline
The detector energy threshold will be significantly lowered if the absolute signal amplitude is increased, without spoiling the baseline detector noise.

\subsection{Light detector performance}
\begin{figure}[h]
\centering
\includegraphics[width=0.4\textwidth]{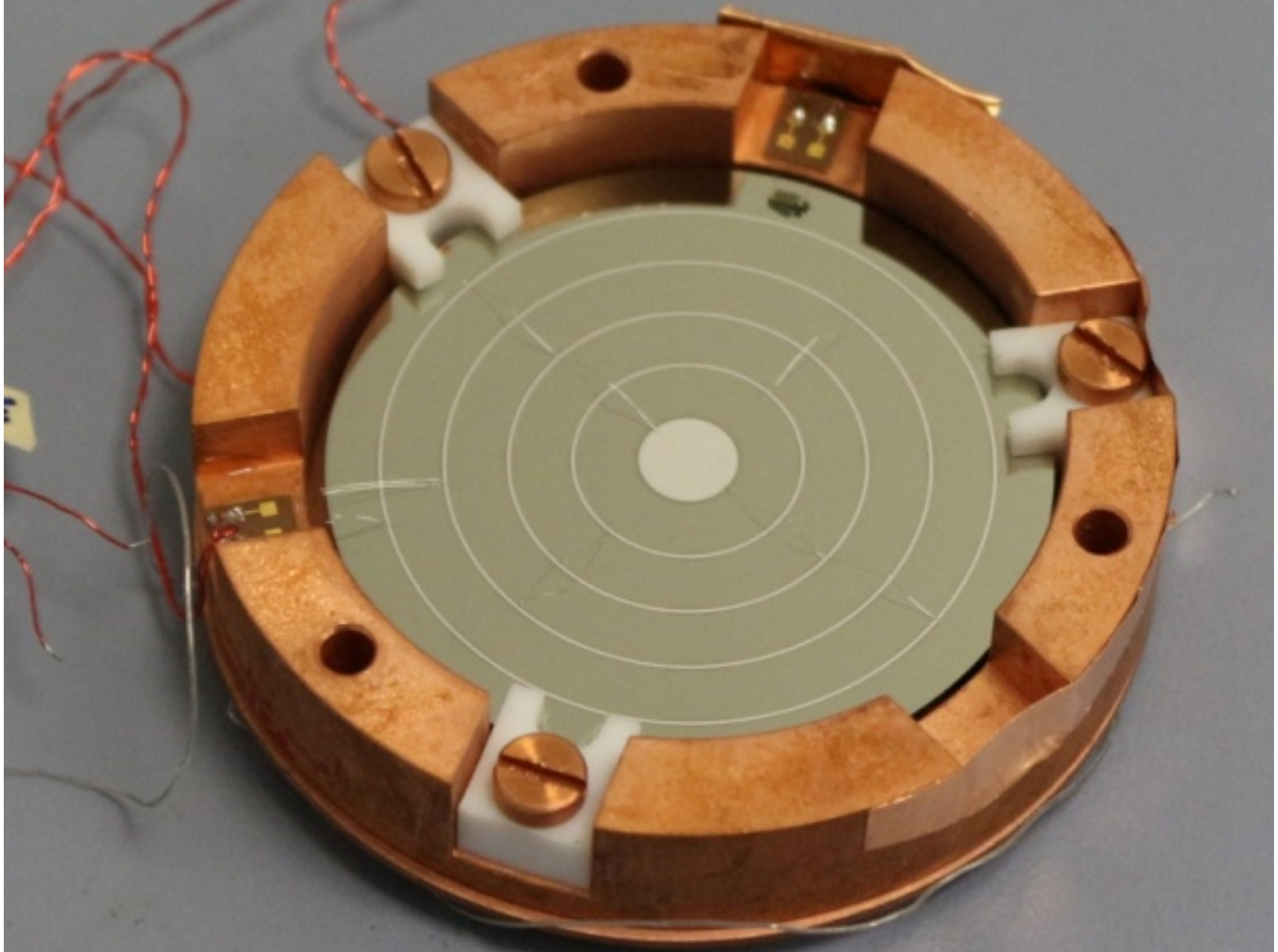}
\caption{\label{fig:LD} Picture of the electrodes deposited on the 5~cm Ge light absorber. The wafer is housed inside a copper structure and kept in position by means of PTFE holders. On the edge of the disk the Ge-NTD sensor is also visible.}
\end{figure}

The detector, produced at CSNSM (Orsay, France), is characterized by aluminum interleaved electrodes, the same design used for the EDELWEISS Dark Matter project~\cite{EDW}, see Fig.~\ref{fig:LD}.

In this work, we report on the performance of a germanium light absorber disk (5~cm of diameter and 180~$\mu$m of thickness) equipped with a Ge-NTD sensor and amplified using the Neganov-Luke effect. The measurements are carried out at the L.N.G.S. underground laboratories of I.N.F.N. (Italy) in the LUCIFER R\&D cryostat~\cite{LUCIFER}.

\begin{figure}[h]
\centering
\includegraphics[width=0.4\textwidth]{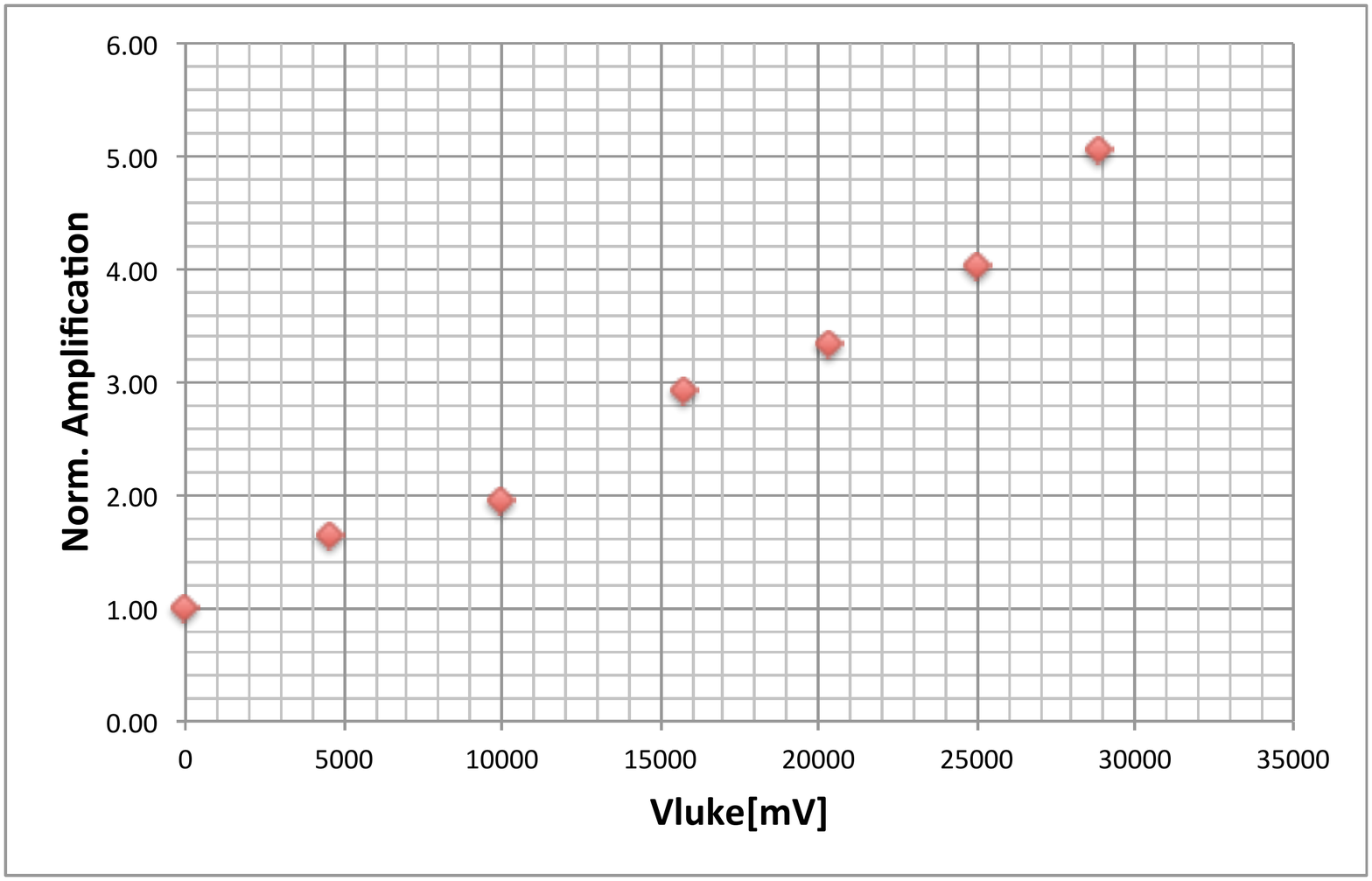}
\includegraphics[width=0.4\textwidth] {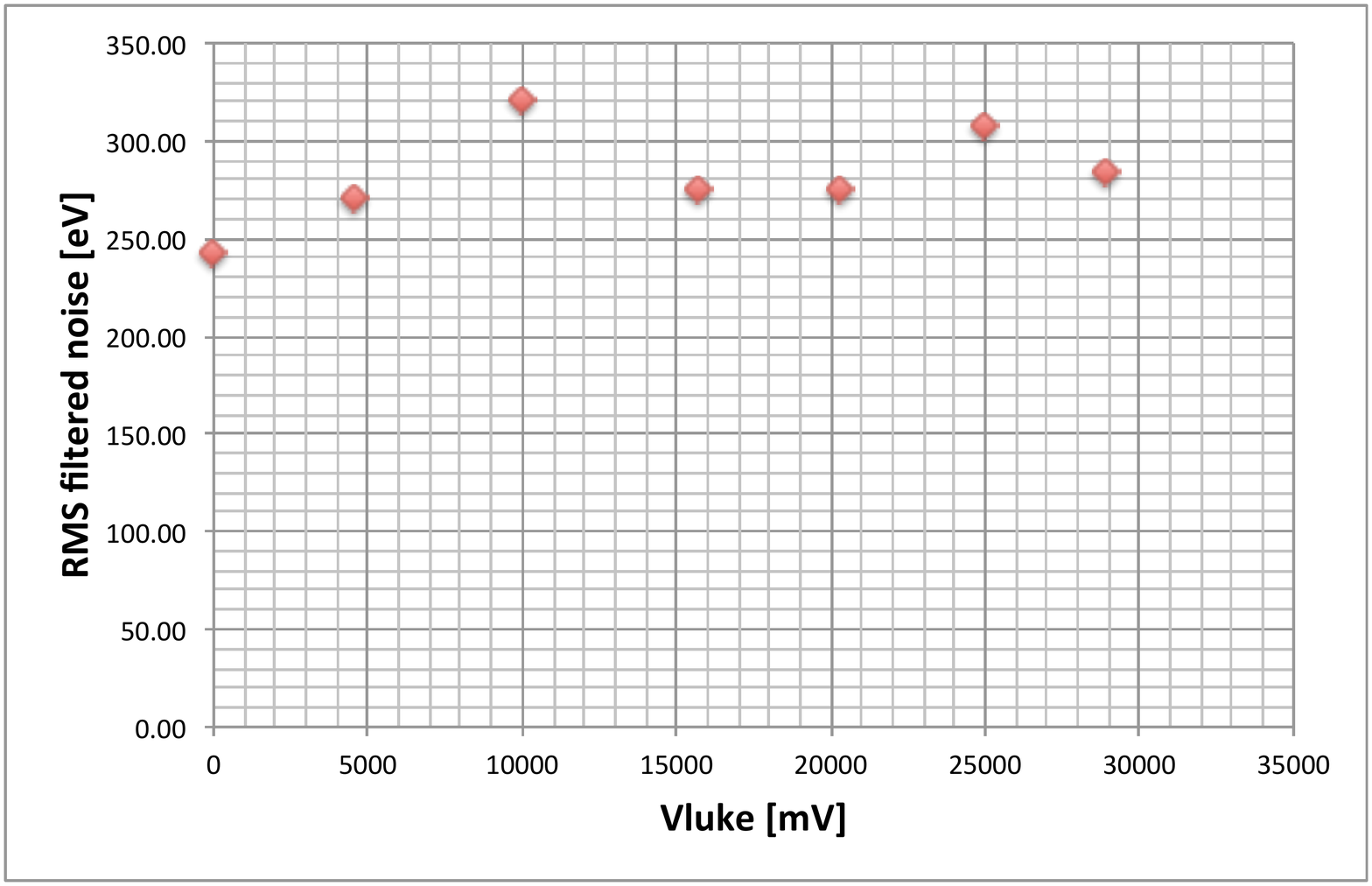}
\caption{\label{fig:perf}Performance of the Neganov-Luke amplification of the light detector: (left) normalized amplification and (right) RMS filtered noise as a function of the voltage applied across the electrodes.}
\end{figure}
The amplification of the detector is studied by means of a LED with a wavelength of 1.45~$\mu$m. Reference pulses are injected on the LD for monitoring the gain change as a function of the voltage applied across the electrodes. In Fig.~\ref{fig:perf}, the performance of the detector: signal amplitude and baseline RMS noise as a function of the voltage are shown.

The electrodes are biased up to a total voltage of 30~V. The detector amplification efficiently works, in fact no relevant worsening of the baseline noise is observed while amplifying the LED thermal signal. The stability of the detector amplification was also tested over long calibration runs and it did not show any spoiling of the detector energy resolution.

\section{Cherenkov light measurement}
Given the good performance of the detector amplification, this was operated faced to a large mass TeO$_2$ CUORE-like bolometer (mass of 750~g). The final goal was to demonstrate the $\alpha$-background suppression by means of the double read-out heat and Cherenkov light. Specifics on the experimental set-up can be found in~\cite{Casali}.

\begin{figure}[h]
\centering
\includegraphics[width=0.45\textwidth]{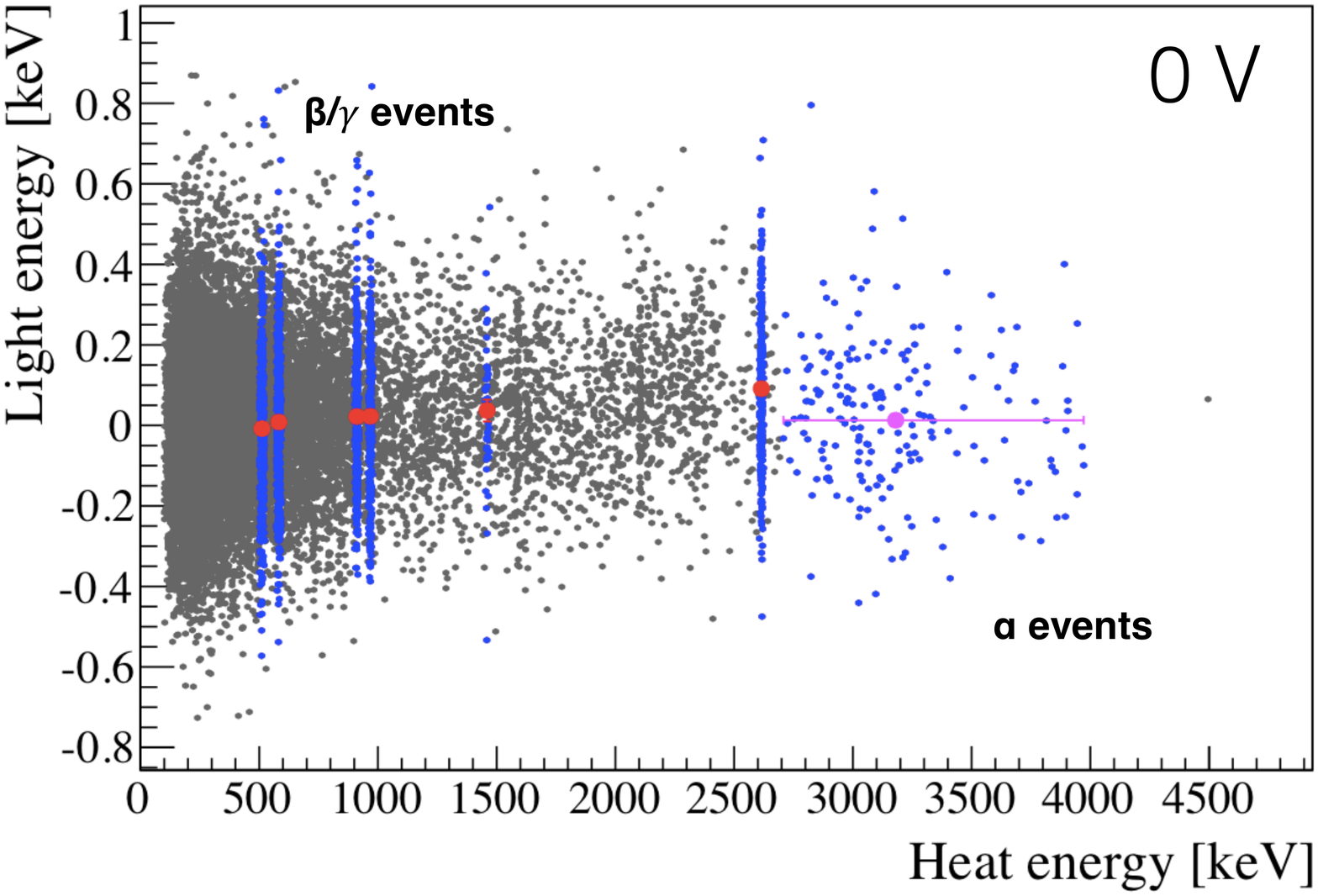}
\includegraphics[width=0.45\textwidth] {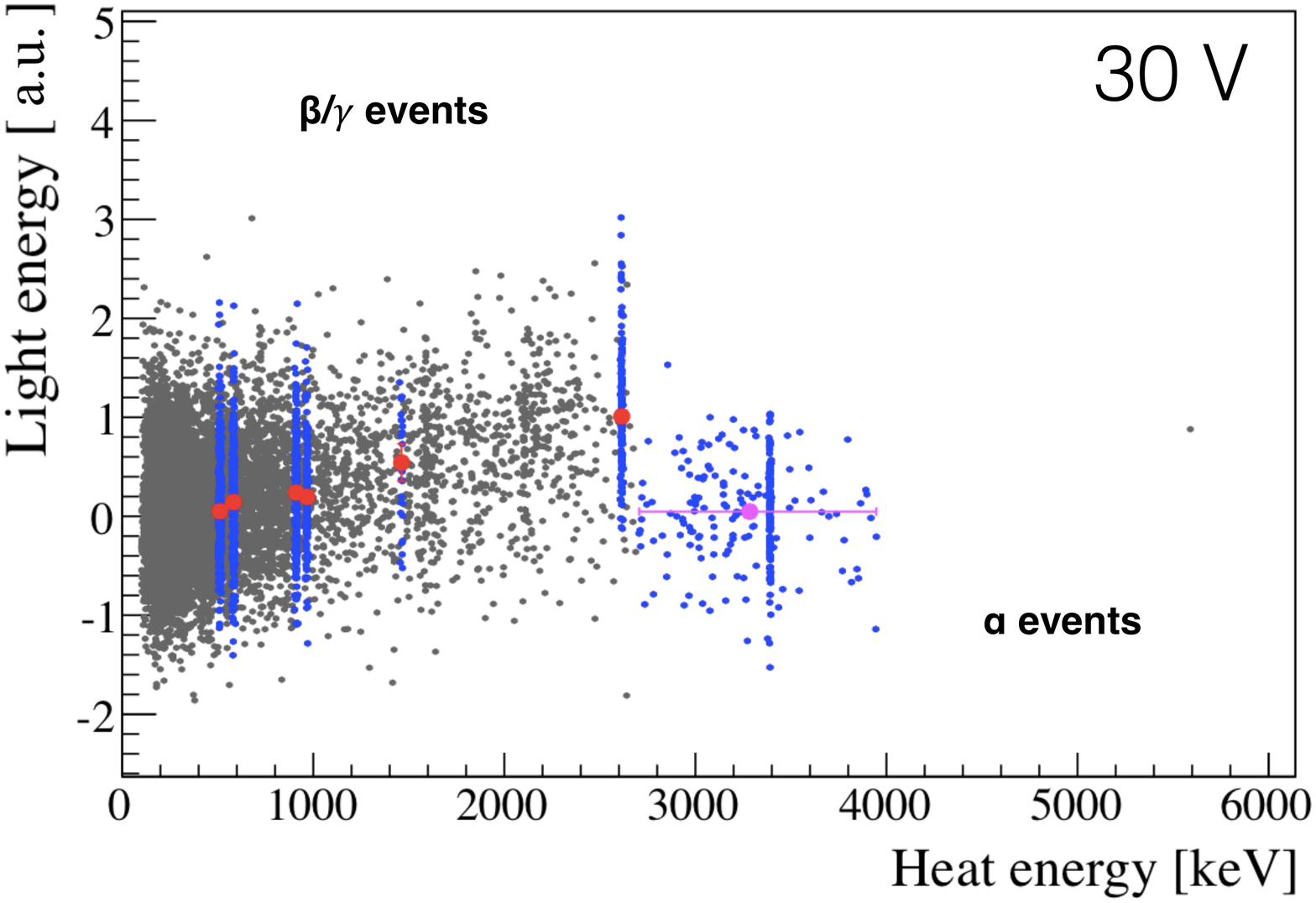}
\includegraphics[width=0.45\textwidth]{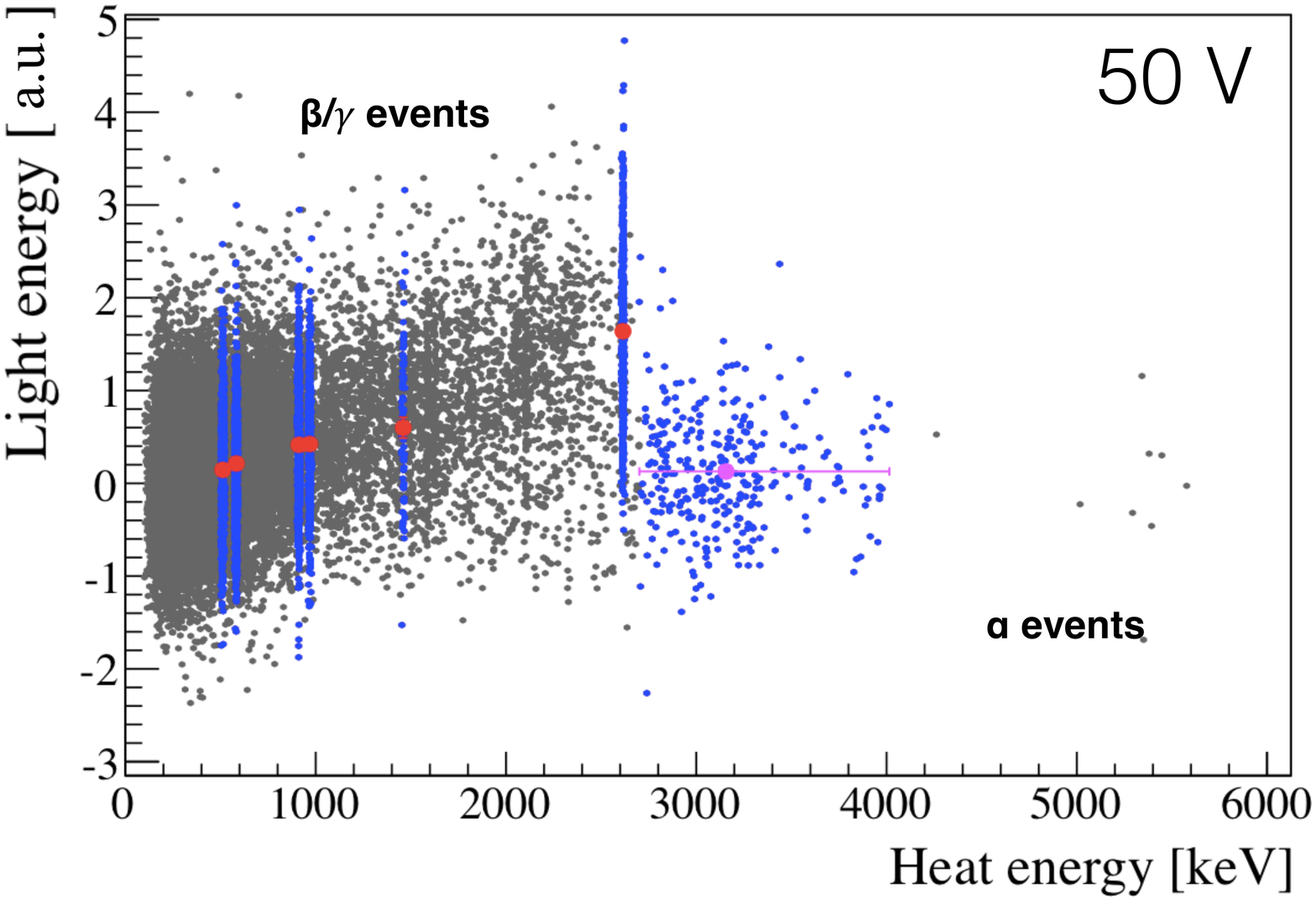}
\includegraphics[width=0.45\textwidth]{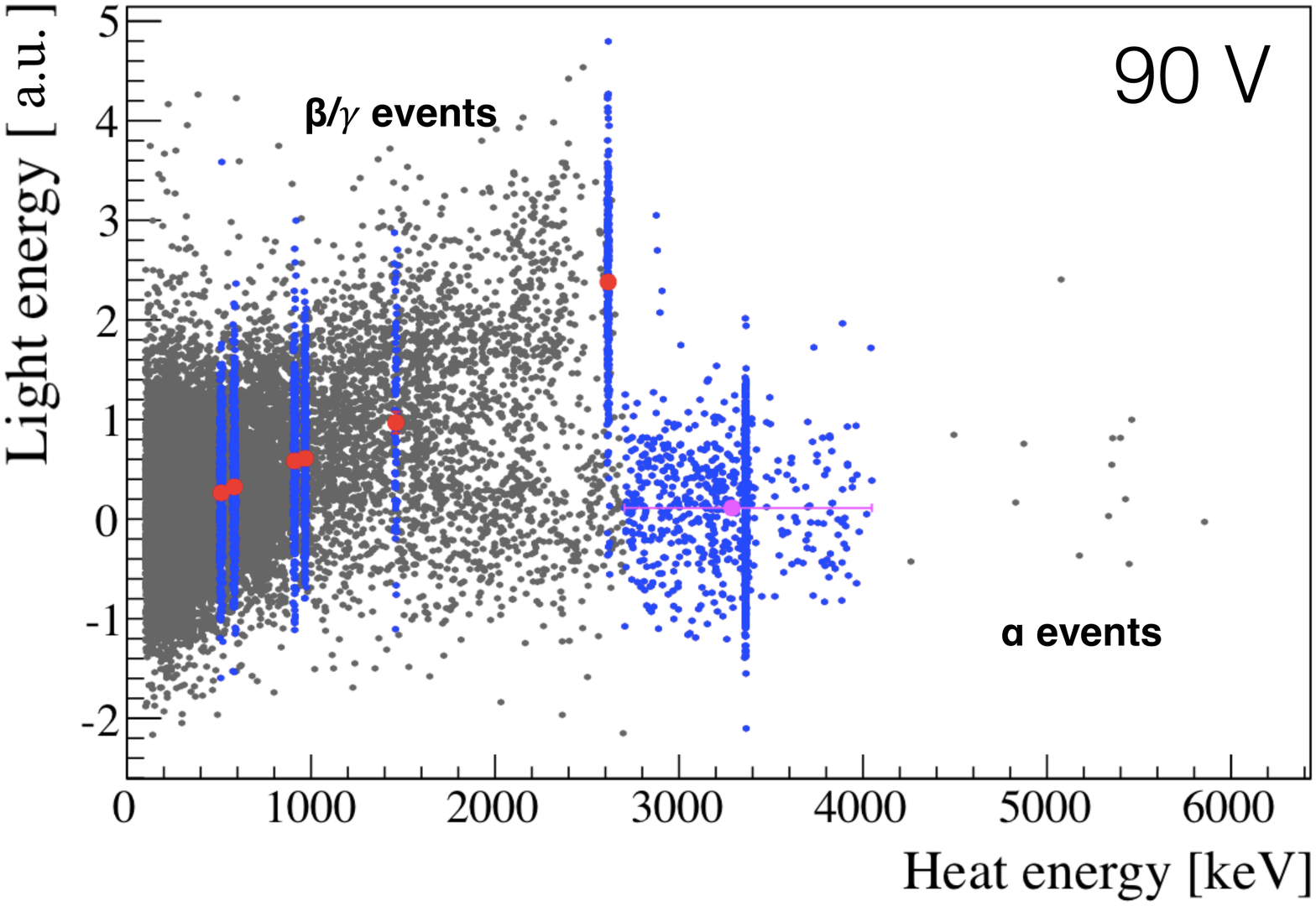}
\caption{Detected light vs. calibrated heat in the TeO$_2$ bolometer for all the acquired events. On the {\it top left} the electric field on the LD is turned off, in the {\it top right} a 30~V bias is applied. In the {\it bottom left} figure the LD is amplified with 50~V of applied voltage, while in the {\it bottom right} 90~V.}
\label{fig:ampl}
\end{figure}

In Fig.~\ref{fig:ampl}, the data for four calibration measurements are shown. The energy scatter plots depicted as detected light versus heat in keV, represent the different configuration of the applied electric field used for the LD amplification. We observe an improvement in the light detector performance, namely the signal amplitude while increasing the bias applied across the detector electrodes. Thus the higher is the intensity of the electric field, the more efficient is the discrimination of the $\alpha$-background in the ROI for $\beta\beta$ investigations.

The amplification of the light signal at 90~V is a factor 9.7 higher than the configuration with no applied field (0~V). If we normalize the detector noise for the amplification gain, the detector RMS baseline noise reduces from 185~eV at 0~V to 19~eV at 90~V. Such low-noise makes an event-by-event identification possible with NTD-based LD, we evaluate a discrimination power between $\alpha$ and $\beta$/$\gamma$ in the ROI of 2.7~$\sigma$, using the same approach described in~\cite{discrimination}. We would like to stress that the absolute signal energy is not changing, since the light source is the same during the entire experiment, but the signal-to-noise ratio is improving.

\section{Conclusions}
The present work shows the potential of Neganov-Luke effect as signal amplitude amplifier for cryogenic light detectors. The developed device ensures high gain without spoiling the baseline noise. Coupling our LD to a CUORE-like massive TeO$_2$ bolometer allows for an efficient particle discrimination to an unprecedented level. Such technology enables the suppression of the $\alpha$-background in the ROI for $^{130}$Te 0$\nu\beta\beta$ decay.



\begin{thebibliography}{99}

\bibitem{CUPID-0}
CUPID Interest Group, ArXiv:1504.03599 (2015).

\bibitem{CUPID-1}
CUPID Interest Group, ArXiv:1504.0361 (2015).

\bibitem{CUORE}
D.R.~Artusa et al., {\it Adv. High Energy Phys.} {\bf 2015}, 879871 (2015).

\bibitem{CUORE0}
K.~Alfonso et al., {\it Phys. Rev. Lett.} {\bf 115}, 102502 (2015).

\bibitem{sticking}
M.~Clemenza et al., {\it Eur. Phys. J. C} {\bf 71}, 1805 (2011).

\bibitem{TTT}
F.~Alessandria et al., {\it Astropart. Phys.} {\bf 45}, 13-22 (2013).

\bibitem{CCVR}
F.~Alessandria et al., {\it Astropart. Phys.} {\bf 35}, 839-849 (2012).

\bibitem{Tabarelli}
T.~Tabarelli~de~Fatis, {\it Eur. Phys. J. C} {65}, 359 (2010).

\bibitem{Casali}
N.~Casali et al., {\it Eur. Phys. J. C} {\bf 75}, 12 (2015).

\bibitem{Schaeffner}
K.~Sch\"affner et al., {\it Astropart. Phys.} {\bf 69}, 30-36 (2015).

\bibitem{TeO2_small}
J.W.~Beeman et al., {\it Astropart.Phys.} {\bf 35}, 558-562 (2012).

\bibitem{MC_light}
F.~Bellini et al., {\it JINST} {\bf 9}, P10014 (2014).

\bibitem{CRESST}
G.~Angloher et al., {\it Eur.Phys.J. C} {\bf 72}, (2012) 1971.

\bibitem{LD_performance}
J.W.~Beeman et al., {\it JINST} {\bf 8}, P07021 (2013).

\bibitem{LUCIFER}
J.W.~Beeman et al., {\it Adv. High Energy Phys.} {\bf2013}, 237973 (2013).

\bibitem{EDW}
E.~Armengaud et al., {\it Phys. Lett. B} {\bf702}, 329-335 (2011).

\bibitem{Luke1}
B.~Neganov,  V.~Trofimov, {\it Otkryt. Izobret.} {\bf 146}, 215 (1985).

\bibitem{Luke2}
P.N.~Luke, {\it J. Appl. Phys.} {\bf 64}, 6858 (1998).

\bibitem{discrimination}
J.W.~Beeman et al., {\it Eur. Phys. J. C} {\bf 72}, 2142 (2012).





\end{thebibliography}
\end{document}